\documentclass[
reprint,
superscriptaddress,
 amsmath,amssymb,
 aps,
prl,
floatfix]{revtex4-2}
\usepackage{xcolor}
\usepackage{graphicx}
\usepackage{dcolumn}
\usepackage{bm}
\usepackage{hyperref}

\begin{document}


\title[]{
Modelling how curved active proteins and shear flow pattern cellular shape and motility\\
}

\author{Shubhadeep Sadhukhan}
\email{shubhadeep.sadhukhan@weizmann.ac.il}
\affiliation{%
 Department of Chemical and Biological Physics, Weizmann Institute of Science, Rehovot, Israel
}
\author{Samo Peni\v{c}}
\email{samo.penic@fe.uni-lj.si}
\affiliation{%
Laboratory of Physics, Faculty of Electrical Engineering, University of Ljubljana, Ljubljana, Slovenia
}
\author{Ale\v{s} Igli\v{c}}
\email{ales.iglic@fe.uni-lj.si}
\affiliation{%
Laboratory of Physics, Faculty of Electrical Engineering, University of Ljubljana, Ljubljana, Slovenia
}
\affiliation{%
Laboratory of Clinical Biophysics, Faculty of Medicine, University of Ljubljana, Ljubljana, Slovenia
}
\author{Nir Gov}%
\email{nir.gov@weizmann.ac.il}
\affiliation{%
 Department of Chemical and Biological Physics, Weizmann Institute of Science, Rehovot, Israel
}
\date{\today}

\begin{abstract}
Cell spreading and motility on an adhesive substrate are driven by the active physical forces generated by the actin cytoskeleton. We have recently shown that coupling curved membrane complexes to protrusive forces, exerted by the actin polymerization that they recruit, provides a mechanism that can give rise to spontaneous membrane shapes and patterns. In the presence of an adhesive substrate, this model was shown to give rise to an emergent motile phenotype, resembling a motile cell. Here, we utilize this ``minimal-cell" model to explore the impact of external shear flow on the cell shape and migration on a uniform adhesive flat substrate. We find that in the presence of shear the motile cell reorients such that its leading edge, where the curved active proteins aggregate, faces the shear flow. The flow-facing configuration is found to minimize the adhesion energy by allowing the cell to spread more efficiently over the substrate. For the non-motile vesicle shapes, we find that they mostly slide and roll with the shear flow. We compare these theoretical results with experimental observations, and suggest that the tendency of many cell types to move against the flow may arise from the very general, and non-cell-type-specific mechanism predicted by our model. 
\end{abstract}

\maketitle

\section{Introduction}
Cell migration plays a crucial role during many key biological processes, from morphogenesis to cancer progression. As a result, the molecular components involved in cell migration, most notably the actin cytoskeleton, have been intensively investigated. Despite the great progress that was made, it is still an open question how do the different cellular components self-organize in a spatial pattern that maintains the robust motile cell shape. Several theoretical models have been proposed to explain the spontaneous emergence of the motile cell shape.

When cells migrate within the blood and lymphatic vessels, they experience fluid flow, which exerts shear forces on the cells. Outstanding examples include lymphocytes in the lymphatic vessels \cite{von2003homing}, neutrophils and  T cells (types of immune cells) rolling and migrating in the blood vessels towards a site of inflamation \cite{luster2005immune,shulman2009lymphocyte}, endothelial cells \cite{alghanem2021swell1} and fibroblasts crawling to sites of injury \cite{kole2005intracellular}. Therefore, an understanding of how shear stresses influence cell movement is essential, and is still lacking. 

One reason for this is that there appear different responses to shear in different cells, and within the same cell type under different conditions, as shown by the following examples. The exposure of sparsely plated endothelial cells, or a wounded monolayer, to shear flow inhibits their migration against the flow ~\cite{ZaidelBar2005}. The direction of T-lymphocyte cell migration under shear flow depends on the adhesion receptors~\cite{Dominguez2015, Anderson2019}: When VCAM-I (Vascular Adhesive Molecule-I) is used, the cells migrate with the flow, while the cells migrate against the flow when ICAM-I (Intracellular adhesive molecule-I) is used. As the shear rate increases, T-lymphocytes favour migration against the flow when ICAM-1 is present, even in the presence of VCAM-1~\cite{Dominguez2015}. The migration of T-cells with the shear also depends on previous exposure to the flow~\cite{Piechocka2021}.

However, one prominent feature that appears consistently in many cell types, is a tendency to migrate up stream against the flow. This behavior was observed in T-lymphocyte cells ~\cite{steiner2010differential,Dominguez2015,Anderson2019}, microvascular endothelial cells ~\cite{Ostrowski2014,alghanem2021swell1}, circulating tumor cells \cite{follain2018hemodynamic}, and in the single-celled amoeba \emph{Dictyostelium discoideum} ~\cite{decave2003shear,Fache2005,Dalous2008}. The origin of this prevalent migration response to shear flow is not understood at present.

Here, we utilize a recently developed theoretical model to explore the response of adherent cells to shear flow. The coarse-grained theoretical model describes the shape dynamics of a vesicle that contains curved membrane proteins that recruit active protrusive forces from the cytoskeleton. This model was shown to give rise to spontaneous pattern formation on the the membrane, resulting in different shapes of vesicle. In the presence of adhesion to an external substrate, a motile phenotype emerges in this model. Here we exert on this vesicle an external force field that emulates the viscous drag force due to the fluid flow. This is an approximate description, that avoids solving the full flow field around the cell, but may provide us with a qualitative understanding of the main physical effects of the shear forces due to the flow. The simplicity of the model makes the calculated results very general, not cell-type-specific. They may therefore shed light on basic physical processes that apply to many cell types, such as the observed tendency of many types of motile cells to migrate upstream.

\begin{figure}
    \centering
    \includegraphics[scale=0.25]{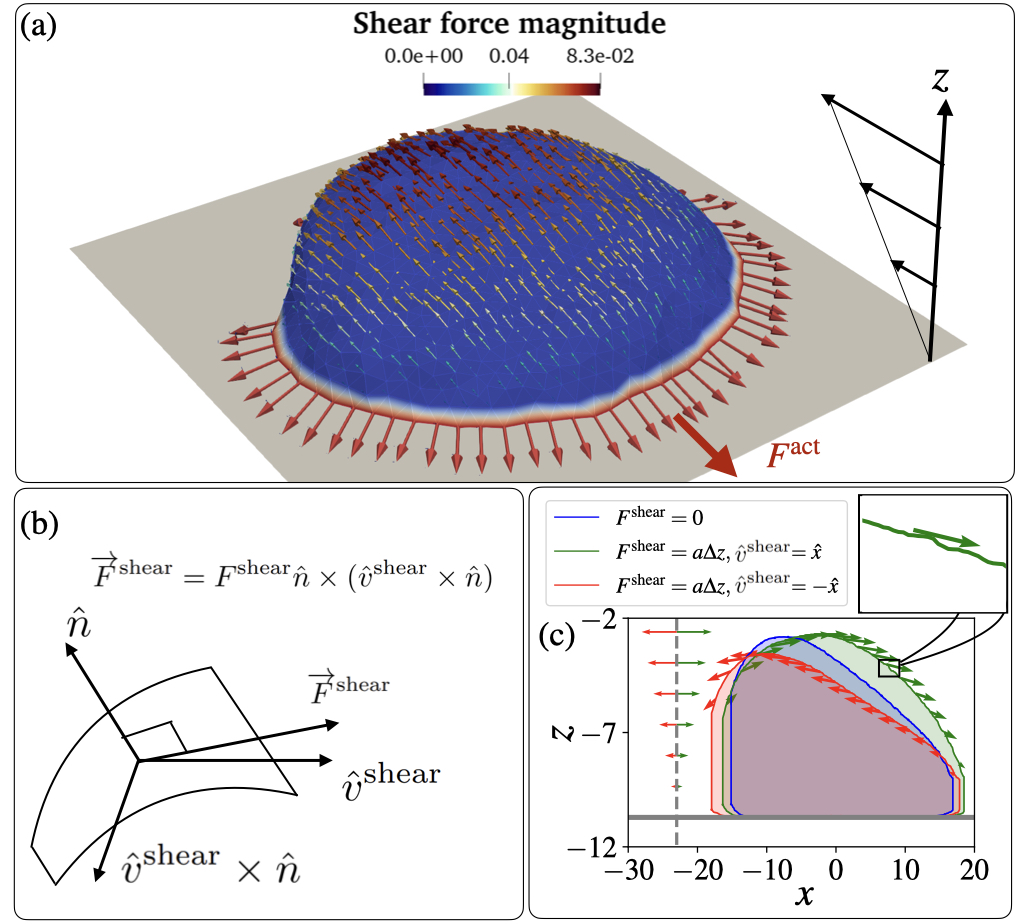}
    \caption{Schematic diagram for a vesicle on an adhesive substrate under active protein forces and the shear force. (a) A highly polarized vesicle is placed on the adhesive surface at $z=z_{\rm ad}$ facing a shear flow. The curved proteins are shown in red colour on the blue-coloured membrane. The parameters used in the simulations are: $F=2 k_BT/l_{min}$, $E_{\rm ad}=3 k_BT$.
    (b) The schematic diagram for the direction of the shear force that is tangential to the vesicle membrane. The direction is calculated as $\hat{n}\times(\hat{v}^{\rm shear}\times\hat{n})$, where, $\hat{n}$ and $\hat{v}^{\rm shear}$ are the normal direction to the surface and the far flow field. The shear force magnitude is given by the linear relation: $F^{\rm shear}=a(z-z_{\rm ad})=a\Delta z$, with $a$ a parameter that determines the magnitude of the shear force.
    (c) The effect of the shear force is demonstrated. The cross sections of the vesicle at $y=y_{\rm avg}$ are shown under three cases. No shear, shear in the positive $x$-direction, and shear in the negative $x$-direction cases are shown in blue, green, and red. Respective arrows show the magnitude and direction of the shear force. The ratio between the maximum shear force on a node and the active force on a node is $0.04$. The ratio between the total shear and active force is approximately $0.7$, when we set $a=0.01k_BT/l_{min}^2$.}
    \label{fig1}
\end{figure}
\section{Model} 
Our model is based on Monte Carlo (MC) simulations to evolve the shape of a vesicle in time (see Supplementary Material). The vesicle is described by a three-dimensional surface (Fig.~\ref{fig1}a) of $N$ vertices, each connected with bonds of length $l$, to form a closed, dynamically triangulated, self-avoiding network, with the topology of a sphere~\cite{Fonari2019}.
The position vector of the $i$th vertex is $\overrightarrow{r_i}$, where $i\in[1, N]$. The vesicle contains mobile curved membrane complexes, which are also sites of force application, representing the protrusive force exerted by actin polymerization. The vesicle is placed on a flat adhesive surface parallel to the $x$-$y$ plane at $z=z_{\rm ad}$. 

The total energy of the vesicle is the sum of various contributions \cite{sadhu2021modelling}: (a) the local bending energy due to the membrane curvature, (b) the energy due to binding between nearest-neighbour membrane protein complexes, (c) the energy due to the active cytoskeleton force, (d) the adhesive energy due to the attractive interaction between the vesicle and the substrate and (e) the energy due to the force experienced by the vesicle due to shear flow.

The bending energy is given by the Helfrich expression~\cite{Helfrich1973} as
\begin{equation}
    W_b=\frac{\kappa}{2}\int_A (C_1+C_2-C_0)^2 dA
    \label{eq:bending_energy}
\end{equation}
where $C_1$ and $C_2$ are principle curvatures, and $\kappa=20k_{B}T$ is the bending rigidity. We set the spontaneous curvature $C_0= 1 l_{min}^{-1}$ for the nodes that contain the curved protein complexes, while it is set to zero for the bare membrane. The interaction energy between nearest-neighbour proteins is expressed as
\begin{equation}
    W_d=-\sum_{i<j} w\mathcal{H}(r_0-r_{ij})
    \label{eq:binding_energy}
\end{equation}

where $\mathcal{H}$ is the Heaviside step function, $r_0$ is the interaction range, $r_{ij}=|\overrightarrow{r_i}-\overrightarrow{r_j}|$ is the distance between proteins, and $w=1~k_BT$ is the interaction energy between neighboring proteins in all the simulations in this paper.

The energy (work) due to the active protrusive force exerted by actin polymerization at the positions of the curved protein complexes

\begin{equation}
    \delta W_F=-F\sum_i \hat{n_i}\cdot \delta\overrightarrow{r_i},
    \label{eq:active_energy}
\end{equation}
where, $\hat{n_i}$ is the outward normal to the membrane and index $i$ runs over the positions of all proteins, $F$ is the strength of the active force, and $\delta\overrightarrow{r_i}$ is the MC shift in the position of the node. The total active force is denoted by $F^{\rm act}$ (Fig.~\ref{fig1}a)

The vesicle can adhere to the adhesive surface located at $z=z_{\rm ad}$, and this energy contribution is
\begin{equation}
    W_{\rm ad}=-\int_A V(z) dA,
    \label{eq:adhesive_energy}
\end{equation}
where $V(z)$ is the interaction potential between the adhesive surface and the vesicle. If the node is close to the surface, $z_{\rm ad}\leq z_i \leq z_{\rm ad}+\delta z$, then the adhesion energy is $V(z)=E_{\rm ad}$, while it is zero for all other nodes. We set $\delta z=l_{\rm min}$ for this whole paper, which is the minimal permitted bond length (to prevent pathological triangulation). The adhesive surface acts as a rigid barrier that the membrane can not penetrate.

Within this model we do not explicitly describe the fluid surrounding and within the vesicle. The MC calculation does not describe the correct time-scale of shape changes, as the dissipative processes involving the fluid flow are not included. This model can predict the shape changes of the vesicle as it minimizes the energy terms listed above. 

This limitation means that when we wish to add the effects of shear forces due to fluid flow, we have to implement some approximate way for exerting these forces. We consider a fluid flow that has a far-field linear profile close to the surface on which the vesicle is adhered (Fig.~\ref{fig1}a), in the direction $\hat{v}^{\rm shear}$. We do not solve the exact flow field around the vesicle, but we assume that the force exerted on the vertices of the vesicle by the flow is everywhere tangential to the vesicle surface, i.e. in the direction of the projection of the shear flow direction on the local tangent plane $(\hat{n}\times (\hat{v}^{\rm shear}\times \hat{n}))$, where $\hat{n}$ is the local outwards normal to the surface (Fig.~\ref{fig1}b). The force on the vertex due to the shear flow is given by: $\overrightarrow{F}^{\rm shear}=F^{\rm shear} (\hat{n}\times (\hat{v}^{\rm shear}\times \hat{n}))$, where the force magnitude due to the shear is assumed to be given by the linear far-field flow velocity at the corresponding distance from the adhesive surface:  $F^{\rm shear}=a(z-z_{\rm ad})=a\Delta z$ (Fig.~\ref{fig1}a,c).

This force due to the shear flow is applied on the nodes of the vesicle as an external force, which gives the following contribution to the energy (work) of the system due to each MC node move (similar to Eq.\ref{eq:active_energy})
\begin{equation}
    \delta W_{s}=-a\sum_i(z_i-z_{\rm ad})~(\hat{n}\times(\hat{v}^{\rm shear}\times \hat{n}))\cdot \delta\overrightarrow{r_i}.
    \label{eq:shear_energy}
\end{equation}
Note that the shear force is applied to each node along the local tangent, which fluctuates due to local membrane shape undulations (inset of Fig.~\ref{fig1}c).

The total energy change of the system, per MC mode, is given by
\begin{equation}
     \delta W=W_b+W_d+ \delta W_F+W_{\rm ad}+ \delta W_s.
\end{equation}

We first verified that our implementation of the effective force due to shear flow produces reasonable results, by calculating the cross-sectional shape of a protein-free vesicle. We found that the vesicle deforms, and tends to lift and detach from the adhesive surface, as the shear flow increases (Fig.S1), similar to the results of previous experimental and theoretical work ~\cite{Cantat1999a}. Adding passive curved proteins can enhance the adhesion of the vesicle, mitigating the tendency of the vesicle to lift and detach due to the shear flow (Fig.S1). Note that similar detachments were also observed for cells exposed to strong shear flows \cite{Decave2002}. We next investigated the response of our vesicle that contains active curved proteins to the shear flow.

\begin{figure*}
    \centering
    \includegraphics[scale=0.35]{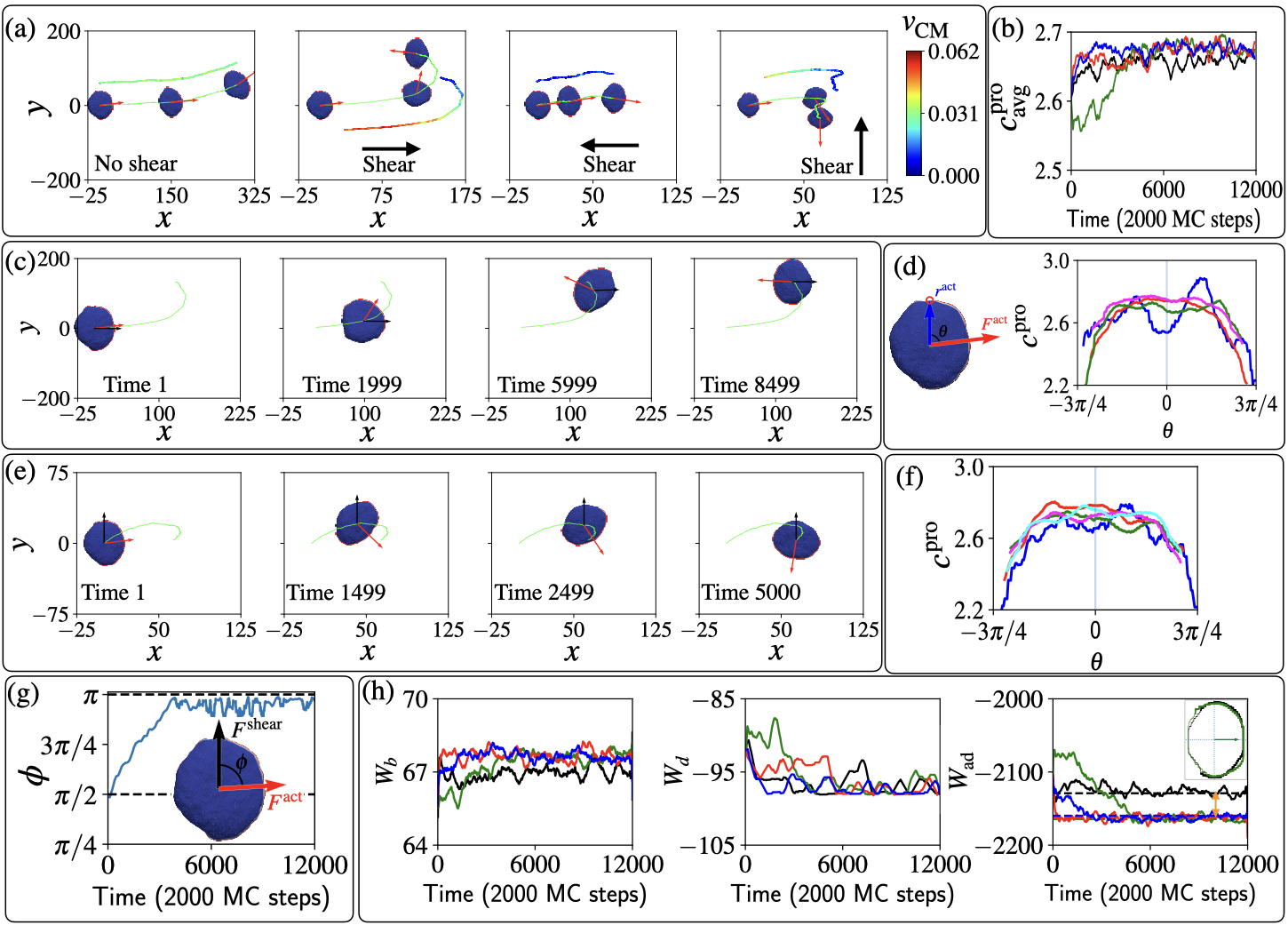}
    \caption{Polarized vesicle under shear flow. (a) The trajectories of the centre of mass of the vesicle are shown in lime colour. The red arrows show the direction of the total active force due to the protein aggregate. The color-coded velocity of the centre of mass is shown a shifted trajectory for each case. (b) The average curvature of the protein aggregate over time for the four shear cases $F^{\rm shear}=0$; $F^{\rm shear}=0.01\Delta z$ along $\hat{v}^{\rm shear}=\hat{x}$, $\hat{v}^{\rm shear}=-\hat{x}$, $\hat{v}^{\rm shear}=\hat{y}$ in black, green, red, and blue solid lines. It shows the drop in the average curvature of the leading edge $c^{\rm pro}_{\rm avg}$ when the shear is parallel to the vesicle's initial polarization $\hat{v}^{\rm shear}=\hat{x}$. (c) Snapshots of the vesicle during the ``U-turn" when the shear is initially parallel to its polarity. (d) The illustration on the left shows the definition of the angle $\theta$ between the position of a protein in the leading-edge cluster with respect to the centre of mass and the direction of the total active force. The right panel shows the curvature profile for the whole protein aggregate at MC times=1, 1999, 4999, 9999 in blue, red, green and magenta solid lines respectively, when $~\hat{v}^{\rm shear}=\hat{x}$. (e) Snapshots of the vesicle during the rotation of the protein aggregate when the shear is initially perpendicular to its polarity. (f) Curvature profile for the whole protein aggregate when shear is initially perpendicular to its polarity at MC times= 1, 499, 2499, 7999, 11499, in blue, red, green, magenta, and cyan lines respectively. (g) The time evolution of the angle $\phi$ between the active force and the shear force. (h) The time evolution of bending energy $W_b$, protein-protein binding energy $W_d$, and adhesion energy $W_{\rm ad}$ for the system shown in (e). The double-headed orange arrow indicates the long-time decrease in adhesion energy when shear is present compared to the case when shear is absent. The inset of the third column shows the adhesive area at MC time=8000, for the sheared and non-sheared cases in green and black lines respectively. We used here: $E_{\rm ad}=3 k_BT$,$F=2k_BT/l_{min}$,$\rho=3.45\%$.}
    \label{fig:motile}
\end{figure*}

\begin{figure}
    \centering
    \includegraphics[scale=0.24]{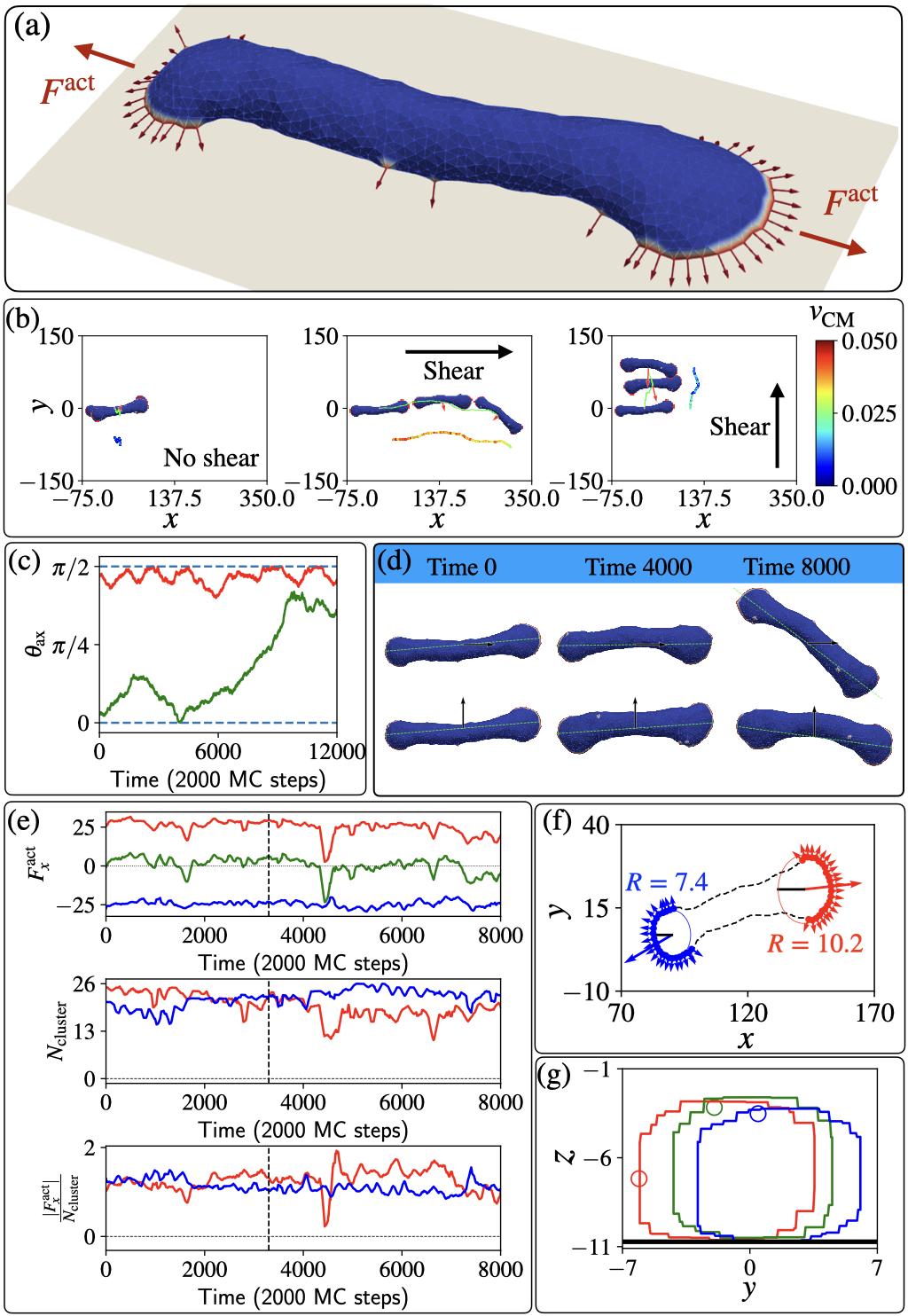}
    \caption{Un-polarized, two-arc vesicle in shear flow. (a) The two-arc, non-motile vesicle, where two opposing leading-edge clusters pull the membrane in the middle into a tubular shape. (b) The trajectories of the centre of mass of the vesicle are shown in lime colour. Red arrows denote the total active force. The velocity of the centre of mass is shown with a colour map by a shifted trajectory for each case in three different columns. (c) Time evolution of the angle $\theta_{\rm ax}$ between the long axis of the vesicle and the shear force when the shear flow is initially parallel and perpendicular to its body axis, in green and red colour respectively. (d) Different shapes of the vesicle over a long time and their body axis for the cases of shear along the body axis and perpendicular to the body axis. (e) In the top line, the active force along $x$ direction for the two different arcs is plotted as function of MC time, in red and blue lines respectively. The green line shows the total active force on the vesicle. In the second line, the numbers of proteins (size of arcs) in the two leading-edge clusters are plotted as function of MC time. In the third line, the time evolution of the efficiency of the arc is quantified by the magnitude of force along $x$ axis per protein in the leading-edge cluster, i.e. $|F^{\rm act}_x|/N_{\rm cluster}$. (f) The projection of positions of the proteins in the two leading-edge arcs on the $x-y$ plane. The circle fit gives the radius of curvature of the corresponding arcs. (g) Rolling motion of the membrane due to shear: The cross-section at the middle of the two-arc-shaped vesicle on the $y$-$z$ plane. We plot at three different times $0, ~50, ~100$ (in unit of $2000$ MC steps) in red, green, and blue respectively. A particular vertex at different times is highlighted with a circular marker to illustrate the rolling motion. We used here: $E_{\rm ad}=1 k_BT,~F=3 k_BT/l_{\rm min},\rho=3.45\%$.}
    \label{fig:nonmotile}
\end{figure}
\section{Motile vesicles}
We have previously found that our minimal-cell model can describe a variety of steady-state shapes of the adhered vesicle \cite{sadhu2021modelling}. One such phenotype is a motile, crescent-shaped vesicle, which appears for strong adhesive interaction $E_{\rm ad}=3k_BT$ and sufficiently large active protrusive forces. Such a motile vesicle has a direction of polarity, determined by the direction of the total active forces ($F^{\rm act}$), due to the local forces applied by the curved proteins that form the leading edge cluster (Fig.~\ref{fig1}a). This shape is self-sustaining, with the highly curved leading edge maintained by the active forces, thereby stabilizing the cluster of curved proteins that seek to minimize their bending energy at such high curvature. 

As shown in Fig.~\ref{fig1}c, when the shear flow acts opposite to the polarity of the vesicle, the two opposing forces tend to deform the vesicle and sharpen its leading edge. This should therefore further stabilize the leading edge cluster of curved proteins. On the contrary, when the shear is in the direction of the polarity, the leading edge becomes less curved, thereby destabilizing the leading edge cluster, as it can not maintain the high curvature that minimizes the bending energy of the curved proteins. We therefore expect that the motile vesicle will respond and modify its motility due the shear forces.

In Fig.~\ref{fig:motile}(a) we show the effects of different shear directions on the trajectories of the motile vesicle. If shear is absent, the motile vesicle moves persistently along its polarity direction, which meanders over time due to random fluctuations. When shear is parallel to the polarity, we find that the protein aggregate reorients and with it the migration path of the vesicle makes a U-turn, to end up facing the shear flow direction. If the shear is opposite to the polarity, the speed is greatly diminished due to the competition between the active force and the shear force, but the migration direction is highly stable, maintaining the upstream path. Finally, when the shear flow is perpendicular to the polarity, we find again that the protein aggregate rotates and reorients to face the shear. 

To explain the origin of these responses of the migration to the shear, we investigate how the shear forces modify the shape of the vesicle (Fig.~\ref{fig1}c), and how these shape changes affect the leading-edge protein aggregate. The average mean-curvature $c^{\rm pro}_{\rm avg}$ of all the vertices with the curved proteins is shown in Fig~\ref{fig:motile}(b). We can see a significant drop in the curvature of the leading edge at the early times for the case when the shear is initially parallel to the polarity direction of the vesicle (green solid line). This is a quantification of the effect shown in Fig.~\ref{fig1}c. As the vesicle reorients to face the shear flow, the average mean-curvature $c^{\rm pro}_{\rm avg}$ increases, and at long times, when the vesicle faces the shear, it is slightly higher in the presence of shear compared to the no-shear case. The proteins aggregate at the leading edge of the vesicle, and prefer the configuration with the higher curvature, which is oriented against the shear flow. 

We follow this reorientation process in Fig.~\ref{fig:motile}(c), and in Fig.~\ref{fig:motile}(d) we plot the local mean curvature
of the proteins along the leading edge ($c^{\rm pro}$). The initial reduction of the curvature due to the shear is most significant in the direction of the shear ($\theta\approx 0$, blue line), as the vesicle is pushed from behind and the front fattens. As time evolves, the proteins rotate and the curvature along the leading edge increases, first in the direction facing the flow (red line, compare negative vs positive angles). Finally, the protein aggregate orients in the direction facing the shear flow, and the middle of the protein aggregate has the highest curvature.

A very similar dynamics is observed for shear that is perpendicular to the initial vesicle polarization, as shown in Fig.~\ref{fig:motile}(e). The vesicle experiences a shear force that pushes from one side, which makes the farthest side of the protein aggregate fatter. Therefore, the protein aggregate rotates towards the more highly curved region and becomes motile against the shear. The curvature at the sites of the proteins $c^{\rm pro}$ is shown in Fig.~\ref{fig:motile}(f). Initially the far side ($\theta \approx \pi/2$) of the vesicle gets fattened, and has lower curvature than the curvature at the side that faces the shear. This gradient in curvature induces the rotation of the leading edge cluster, as shown in Fig.~\ref{fig:motile}(g).

The dynamics in our model is driven by minimization of energy and work. In Fig~\ref{fig:motile}(h) we plot the time evolution of the different energy components: the bending energy $W_b$, the binding energy $W_d$, and the adhesion energy $W_{\rm ad}$. At the steady-state configurations, when the vesicle is polarized against the flow, we find that the bending energy is increased due to shear, compared to the case when shear is absent. So clearly this energy is not minimized during the reorientation process. The protein-protein binding energy at short times is largest when the leading edge cluster is destabilized by the parallel shear flow (green line), but at long times this energy is essentially unaffected by the presence of shear.

Finally, we find that the adhesion energy $W_{\rm ad}$ is clearly smaller (more negative) at long times in the presence of shear compared to the no-shear case, as indicated by the two-headed orange arrow.
We plot the adhesive area of the vesicle that is in contact with the substrate, for the case with and without shear, in the inset of the rightmost panel of Fig~\ref{fig:motile}(h), showing the bigger adhesive area when the shear is present. 
At a lower shear flow parameter, we found that the results are qualitatively the same, but the reorientation dynamics take longer to occur.

We therefore conclude that our simplified, minimal-cell model can provide a physical mechanism for the stabilization of cell migration that is upstream in the presence of shear flow. The basis for this mechanism is the increased cell spreading due to shear flow (Fig~\ref{fig:motile}(h)), which was also observed in cells~\cite{Dominguez2015, ChotardGhodsnia2007}. Note that cells respond to shear also through signalling that modify the overall cell behavior, which corresponds in our model to changes to the model parameters. Nevertheless, the physical mechanism that we find for migration against the flow is not cell-type-specific and is independent of any complex biochemical signalling.
It may therefore explain why this behavior appears in many different cell types, as listed in the introduction.


\section{Non-motile vesicles}
Next, we explore the response to shear flow for non-motile adhered vesicles in our model. The non-polar, non-motile phenotypes of adhered vesicles in our model have several shapes \cite{sadhu2021modelling}: At low adhesion or low active force, the vesicle is weakly spread and has a roughly hemispherical shape. For high concentration of the curved membrane proteins, and at sufficiently high adhesion or active force, the vesicle spreads into a round pancake-like shape with a closed, circular leading-edge. The response of both of these shapes to the shear is given in the SI (Fig.S2,S3), where its shown that they roll and slide with the flow.

A more interesting shape arises on surfaces with weaker adhesion, or at high active forces, where the vesicle spreads into a two-arc shape (Fig~\ref{fig:nonmotile}a). The vesicle is elongated by two leading-edge clusters at opposing ends, with the membrane between them being pulled into a cylindrical shape. Adhered cells often have such elongated shapes, with multiple, competing leading edge lamellipodia, which render them non-polar and non-motile \cite{pankov2005rac,schaufler2016selective,singh2020cell,dimchev2021induced}.


In Fig~\ref{fig:nonmotile}(b) we plot the trajectories of the two-arc vesicle for three different shear conditions with respect to the initial long axis of the vesicle (the long axis is calculated as explain in the SI). When shear is absent, the vesicle is almost completely non-motile, while in the presence of flow the vesicle moves with the shear flow. Interestingly, we can see that initially the vesicle moves faster when the shear is in the same direction as its body axis, compared to when the shear is perpendicular to the vesicle's long axis. However, this migration along the long axis is unstable, and at long times the vesicle rotates to being perpendicular to the flow (Fig~\ref{fig:nonmotile}c), which is the stable configuration. In Fig~\ref{fig:nonmotile}(d) we show snapshots of the vesicle at different times as it moves with the shear flow, either parallel or perpendicular to the initial long axis of the vesicle. 

As shown in Fig~\ref{fig:nonmotile}(b), the vesicle is moving faster when the shear is along the body axis compared to when the shear is perpendicular to the body axis. This indicates that the shear is inducing some polarization of the active forces, which now have a net force that contributes to the active motility along the shear flow. To understand the origin of this shear-induced polarization, we analyzed the two leading-edge clusters at the opposing ends of the vesicle. We denote the arc pulling towards positive $x$ direction (with the flow) and negative $x$ direction (against the flow), arc-1 (red) and arc-2 (blue) respectively (Fig~\ref{fig:nonmotile}e,f). In Fig~\ref{fig:nonmotile}(e) we show that the net active force due to the two clusters  is positive at the early times, indicating that indeed there is a net active force from the leading edges, pulling the vesicle with the flow direction. This shear-induced asymmetry is manifested as a larger active force along $x$-direction due to arc-1 compared to the negative component from arc-2.
However, the sizes of the two arcs $N_{\rm cluster}$ do not show any systematic difference between the two leading edge clusters. Nevertheless, the net force in the flow direction due to arc-1 is stronger than the force due to arc-2 since the efficiency, defined as the net force along the flow direction per protein $|F^{\rm act}_x|/N_{\rm cluster}$, of the proteins in arc-1 is larger (Fig~\ref{fig:nonmotile}e). 
Even when the size of arc-1 is smaller than arc-2, the proteins in arc-1 can be more efficient and produce a stronger net force along the flow direction, make the vesicle motile in the presence of shear.

To get more insight into this efficiency of the two leading edges, we plotted the positions of the proteins on the $x$-$y$ plane as shown in Fig~\ref{fig:nonmotile}(f), at a time where the sizes of the two clusters is almost identical, yet there is a net force in the direction of arc-1. We fit a circular arc and find the radius of curvature for each leading edge cluster using the gradient-descent method. We find that the radius of curvature is bigger for arc-1 ($R=10.2l_{\rm min}$) compared to arc-2 ($R=7.4l_{\rm min}$), and this flatter shape of arc-1 makes its proteins' active forces more oriented along the flow, compared to the orientations of the active forces in arc-2. The flatter shape of the leading edge of arc-1 is due to the shear forces pushing membrane along the tubular part that connects the two leading edges, such that membrane area is forced from the region of arc-2 to that of arc-1, allowing the fan-shaped region of arc-1 to grow larger in area.

In the stable phase, where the vesicle moves with the shear flow that is perpendicular to its body axis, We plotted the cross-sectional area of the vesicle along its middle point ($x_{\rm avg}$) at different times (Fig~\ref{fig:nonmotile}g). By following one particular node we illustrate that the membrane is rolling on the surface due to the flow-induced shear forces. In the SI we present the dynamics of the two-arc shape at different shear flow strengths.


Weakly polarized cells, such as Chinese hamster ovary (CHO) cells, exhibit weak migration with the shear flow, often maintaining an elogated shape that is perpendicular to the shear direction~\cite{Dikeman2008}. These CHO cells tend to spread in a circularly symmetric manner, and indeed our vesicles that spread uniformly tend to slide with the shear flow (SI, Fig.S3), as observed for these cells.

\section{Conclusions}
Our ``minimal-cell" model, where cell spreading and migration emerges due to curved membrane proteins that recruit the protrusive forces of actin polymerization, is used to explore the effects of shear forces applied to the membrane due to an imposed fluid flow. This model shows that since the self-organization of the curved proteins and active forces are dependent on the membrane shape, the system is strongly affected by these flow-induced shear forces. 

We found that the motile crescent-shaped vesicle in our model spontaneously migrates against the shear flow due to the reorganization of curved protein in response to the shear flow. This behavior arises simply from the physics of minimizing the adhesion energy to the substrate. Since our mechanism is based on a very simple model, and a physical mechanism, it may explain why the tendency of cells to migrate against the flow appears in many different cell types that migrate using lamellipodia protrusions ~\cite{steiner2010differential,Dominguez2015,Anderson2019,Ostrowski2014,alghanem2021swell1,follain2018hemodynamic,decave2003shear,Fache2005,Dalous2008}. Though our model does not include many cellular components, it offers an explanation for the origin of this prevalent migration response to shear flow, which is not understood at present.

For the non-motile vesicles we found that they tend to migrate or roll with the shear flow, which may explain why weakly motile cells tend to move with the flow ~\cite{decave2003shear, Dikeman2008, Fache2005}. Note that cells respond to shear flow due to signalling pathways \cite{rose2007integrin,chistiakov2017effects}, which modify the overall cell-substrate adhesion and cytoskeleton activity. This layer of biochemical control manifests as modifications to the parameters of the vesicle in our model, beyond the shear-induced shape changes that we investigated. 

The MC model we used here does not describe the full fluid-flow field surrounding the cell. Future studies that include explicitly the fluid dynamics \cite{noguchi2004fluid,mauer2018flow} and additional cellular components \cite{dabagh2017mechanotransmission}, may be used to explore the dynamics predicted by our model with better physical realism, at the price of greatly increased complexity and computation time. 

Our results may also explain the motility response of cells to external forces that are exerted on them by other means, not due to fluid flow. In \cite{weber2012mechanoresponsive} it was shown that when an adhered cell is pulled by a magnetic beads that is attached to the cell, it tends to polarize in the opposite direction to the applied force. Similarly, when cells that are attached to each other exert a pulling force on each other, they tend to polarize in opposite directions to each other \cite{weber2012mechanoresponsive}. This plays an important role during collective cell migration \cite{mayor2016front}. These responses may arise from the same behavior that we obtain here, namely the pulling force tends to stabilize the leading edge of the cell in the direction that is opposite to the direction of the external force (Fig.\ref{fig1}).

\section*{Acknowledgement}
We thank Raj Kumar Sadhu and Yoav Ravid for many interesting and fruitful discussions. N.S.G. is the incumbent of the Lee and William Abramowitz Professorial Chair of Biophysics, and acknowledges support by the Ben May Center for Theory and Computation, and the Israel Science Foundation (Grant No. 207/22). This research is made possible in part by the historic generosity of the Harold Perlman Family. A.I. and S.P. were supported by the Slovenian Research Agency (ARRS) through the Grant No. J3-3066 and J2-4447 and Programme No. P2-0232.
\bibliographystyle{abbrv}
\bibliography{shubhadeep_biophys}

\begin{thebibliography}{10}

\bibitem{alghanem2021swell1}
A.~F. Alghanem, J.~Abello, J.~M. Maurer, A.~Kumar, C.~M. Ta, S.~K. Gunasekar,
  U.~Fatima, C.~Kang, L.~Xie, O.~Adeola, et~al.
\newblock The swell1-lrrc8 complex regulates endothelial akt-enos signaling and
  vascular function.
\newblock {\em Elife}, 10:e61313, 2021.

\bibitem{Anderson2019}
N.~R. Anderson, A.~Buffone, and D.~A. Hammer.
\newblock T lymphocytes migrate upstream after completing the leukocyte
  adhesion cascade.
\newblock {\em Cell Adhesion {\&} Migration}, 13(1):164--169, Jan. 2019.

\bibitem{Cantat1999a}
I.~Cantat and C.~Misbah.
\newblock Lift force and dynamical unbinding of adhering vesicles under shear
  flow.
\newblock {\em Phys. Rev. Lett.}, 83:880--883, Jul 1999.

\bibitem{chistiakov2017effects}
D.~A. Chistiakov, A.~N. Orekhov, and Y.~V. Bobryshev.
\newblock Effects of shear stress on endothelial cells: go with the flow.
\newblock {\em Acta physiologica}, 219(2):382--408, 2017.

\bibitem{ChotardGhodsnia2007}
R.~Chotard-Ghodsnia, O.~Haddad, A.~Leyrat, A.~Drochon, C.~Verdier, and
  A.~Duperray.
\newblock Morphological analysis of tumor cell/endothelial cell interactions
  under shear flow.
\newblock {\em Journal of Biomechanics}, 40(2):335--344, Jan. 2007.

\bibitem{dabagh2017mechanotransmission}
M.~Dabagh, P.~Jalali, P.~J. Butler, A.~Randles, and J.~M. Tarbell.
\newblock Mechanotransmission in endothelial cells subjected to oscillatory and
  multi-directional shear flow.
\newblock {\em Journal of The Royal Society Interface}, 14(130):20170185, 2017.

\bibitem{Dalous2008}
J.~Dalous, E.~Burghardt, A.~M\"{u}ller-Taubenberger, F.~Bruckert, G.~Gerisch,
  and T.~Bretschneider.
\newblock Reversal of cell polarity and actin-myosin cytoskeleton
  reorganization under mechanical and chemical stimulation.
\newblock {\em Biophysical Journal}, 94(3):1063--1074, Feb. 2008.

\bibitem{decave2003shear}
E.~D{\'e}cav{\'e}, D.~Rieu, J.~Dalous, S.~Fache, Y.~Br{\'e}chet, B.~Fourcade,
  M.~Satre, and F.~Bruckert.
\newblock Shear flow-induced motility of dictyostelium discoideum cells on
  solid substrate.
\newblock {\em Journal of cell science}, 116(21):4331--4343, 2003.

\bibitem{Dikeman2008}
D.~A. Dikeman, L.~A.~R. Rosado, T.~A. Horn, C.~S. Alves, K.~Konstantopoulos,
  and J.~T. Yang.
\newblock alpha4 beta1-integrin regulates directionally persistent cell
  migration in response to shear flow stimulation.
\newblock {\em American Journal of Physiology-Cell Physiology},
  295(1):C151--C159, July 2008.

\bibitem{dimchev2021induced}
V.~Dimchev, I.~Lahmann, S.~A. Koestler, F.~Kage, G.~Dimchev, A.~Steffen, T.~E.
  Stradal, F.~Vauti, H.-H. Arnold, and K.~Rottner.
\newblock Induced arp2/3 complex depletion increases fmnl2/3 formin expression
  and filopodia formation.
\newblock {\em Frontiers in cell and developmental biology}, 9:634708, 2021.

\bibitem{Dominguez2015}
G.~A. Dominguez, N.~R. Anderson, and D.~A. Hammer.
\newblock The direction of migration of t-lymphocytes under flow depends upon
  which adhesion receptors are engaged.
\newblock {\em Integrative Biology}, 7(3):345--355, 2015.

\bibitem{Decave2002}
E.~Décavé, D.~Garrivier, Y.~Bréchet, B.~Fourcade, and F.~Bruckert.
\newblock Shear flow-induced detachment kinetics of dictyostelium discoideum
  cells from solid substrate.
\newblock {\em Biophysical Journal}, 82(5):2383--2395, 2002.

\bibitem{Fache2005}
S.~Fache, J.~Dalous, M.~Engelund, C.~Hansen, F.~Chamaraux, B.~Fourcade,
  M.~Satre, P.~Devreotes, and F.~Bruckert.
\newblock {Calcium mobilization stimulates Dictyostelium discoideum
  shear-flow-induced cell motility}.
\newblock {\em Journal of Cell Science}, 118(15):3445--3458, 08 2005.

\bibitem{follain2018hemodynamic}
G.~Follain, N.~Osmani, A.~S. Azevedo, G.~Allio, L.~Mercier, M.~A. Karreman,
  G.~Solecki, M.~J.~G. Le{\`o}n, O.~Lefebvre, N.~Fekonja, et~al.
\newblock Hemodynamic forces tune the arrest, adhesion, and extravasation of
  circulating tumor cells.
\newblock {\em Developmental cell}, 45(1):33--52, 2018.

\bibitem{Fonari2019}
M.~Fo{\v{s}}nari{\v{c}}, S.~Peni{\v{c}}, A.~Igli{\v{c}}, V.~Kralj-Igli{\v{c}},
  M.~Drab, and N.~S. Gov.
\newblock Theoretical study of vesicle shapes driven by coupling curved
  proteins and active cytoskeletal forces.
\newblock {\em Soft Matter}, 15(26):5319--5330, 2019.

\bibitem{Helfrich1973}
W.~Helfrich.
\newblock Elastic properties of lipid bilayers: Theory and possible
  experiments.
\newblock {\em Zeitschrift f\"{u}r Naturforschung C}, 28(11-12):693--703, Dec.
  1973.

\bibitem{kole2005intracellular}
T.~P. Kole, Y.~Tseng, I.~Jiang, J.~L. Katz, and D.~Wirtz.
\newblock Intracellular mechanics of migrating fibroblasts.
\newblock {\em Molecular biology of the cell}, 16(1):328--338, 2005.

\bibitem{luster2005immune}
A.~D. Luster, R.~Alon, and U.~H. von Andrian.
\newblock Immune cell migration in inflammation: present and future therapeutic
  targets.
\newblock {\em Nature immunology}, 6(12):1182--1190, 2005.

\bibitem{mauer2018flow}
J.~Mauer, S.~Mendez, L.~Lanotte, F.~Nicoud, M.~Abkarian, G.~Gompper, and D.~A.
  Fedosov.
\newblock Flow-induced transitions of red blood cell shapes under shear.
\newblock {\em Physical review letters}, 121(11):118103, 2018.

\bibitem{mayor2016front}
R.~Mayor and S.~Etienne-Manneville.
\newblock The front and rear of collective cell migration.
\newblock {\em Nature reviews Molecular cell biology}, 17(2):97--109, 2016.

\bibitem{noguchi2004fluid}
H.~Noguchi and G.~Gompper.
\newblock Fluid vesicles with viscous membranes in shear flow.
\newblock {\em Physical review letters}, 93(25):258102, 2004.

\bibitem{Ostrowski2014}
M.~A. Ostrowski, N.~F. Huang, T.~W. Walker, T.~Verwijlen, C.~Poplawski, A.~S.
  Khoo, J.~P. Cooke, G.~G. Fuller, and A.~R. Dunn.
\newblock Microvascular endothelial cells migrate upstream and align against
  the shear stress field created by impinging flow.
\newblock {\em Biophysical Journal}, 106(2):366--374, Jan. 2014.

\bibitem{pankov2005rac}
R.~Pankov, Y.~Endo, S.~Even-Ram, M.~Araki, K.~Clark, E.~Cukierman,
  K.~Matsumoto, and K.~M. Yamada.
\newblock A rac switch regulates random versus directionally persistent cell
  migration.
\newblock {\em The Journal of cell biology}, 170(5):793--802, 2005.

\bibitem{Piechocka2021}
I.~K. Piechocka, S.~Keary, A.~Sosa-Costa, L.~Lau, N.~Mohan, J.~Stanisavljevic,
  K.~J. Borgman, M.~Lakadamyali, C.~Manzo, and M.~F. Garcia-Parajo.
\newblock Shear forces induce {ICAM}-1 nanoclustering on endothelial cells that
  impact on t-cell migration.
\newblock {\em Biophysical Journal}, 120(13):2644--2656, July 2021.

\bibitem{rose2007integrin}
D.~M. Rose, R.~Alon, and M.~H. Ginsberg.
\newblock Integrin modulation and signaling in leukocyte adhesion and
  migration.
\newblock {\em Immunological reviews}, 218(1):126--134, 2007.

\bibitem{sadhu2021modelling}
R.~K. Sadhu, S.~Peni{\v{c}}, A.~Igli{\v{c}}, and N.~S. Gov.
\newblock Modelling cellular spreading and emergence of motility in the
  presence of curved membrane proteins and active cytoskeleton forces.
\newblock {\em The European Physical Journal Plus}, 136(5):495, 2021.

\bibitem{schaufler2016selective}
V.~Schaufler, H.~Czichos-Medda, V.~Hirschfeld-Warnecken, S.~Neubauer,
  F.~Rechenmacher, R.~Medda, H.~Kessler, B.~Geiger, J.~P. Spatz, and E.~A.
  Cavalcanti-Adam.
\newblock Selective binding and lateral clustering of $\alpha$ 5 $\beta$ 1 and
  $\alpha$ v $\beta$ 3 integrins: Unraveling the spatial requirements for cell
  spreading and focal adhesion assembly.
\newblock {\em Cell adhesion \& migration}, 10(5):505--515, 2016.

\bibitem{shulman2009lymphocyte}
Z.~Shulman, V.~Shinder, E.~Klein, V.~Grabovsky, O.~Yeger, E.~Geron,
  A.~Montresor, M.~Bolomini-Vittori, S.~W. Feigelson, T.~Kirchhausen, et~al.
\newblock Lymphocyte crawling and transendothelial migration require chemokine
  triggering of high-affinity lfa-1 integrin.
\newblock {\em Immunity}, 30(3):384--396, 2009.

\bibitem{singh2020cell}
S.~P. Singh, P.~A. Thomason, S.~Lilla, M.~Schaks, Q.~Tang, B.~L. Goode, L.~M.
  Machesky, K.~Rottner, and R.~H. Insall.
\newblock Cell--substrate adhesion drives scar/wave activation and
  phosphorylation by a ste20-family kinase, which controls pseudopod lifetime.
\newblock {\em PLoS biology}, 18(8):e3000774, 2020.

\bibitem{steiner2010differential}
O.~Steiner, C.~Coisne, R.~Cecchelli, R.~Boscacci, U.~Deutsch, B.~Engelhardt,
  and R.~Lyck.
\newblock Differential roles for endothelial icam-1, icam-2, and vcam-1 in
  shear-resistant t cell arrest, polarization, and directed crawling on
  blood--brain barrier endothelium.
\newblock {\em The Journal of Immunology}, 185(8):4846--4855, 2010.

\bibitem{von2003homing}
U.~H. von Andrian and T.~R. Mempel.
\newblock Homing and cellular traffic in lymph nodes.
\newblock {\em Nature Reviews Immunology}, 3(11):867--878, 2003.

\bibitem{weber2012mechanoresponsive}
G.~F. Weber, M.~A. Bjerke, and D.~W. DeSimone.
\newblock A mechanoresponsive cadherin-keratin complex directs polarized
  protrusive behavior and collective cell migration.
\newblock {\em Developmental cell}, 22(1):104--115, 2012.

\bibitem{ZaidelBar2005}
R.~Zaidel-Bar, Z.~Kam, and B.~Geiger.
\newblock Polarized downregulation of the paxillin-p130cas-rac1 pathway induced
  by shear flow.
\newblock {\em Journal of Cell Science}, 118(17):3997--4007, Sept. 2005.

\end{thebibliography}


\begin{thebibliography}{1}

\bibitem{Cantat1999a}
I.~Cantat and C.~Misbah.
\newblock Lift force and dynamical unbinding of adhering vesicles under shear
  flow.
\newblock {\em Phys. Rev. Lett.}, 83:880--883, Jul 1999.

\bibitem{Fonari2019}
M.~Fo{\v{s}}nari{\v{c}}, S.~Peni{\v{c}}, A.~Igli{\v{c}}, V.~Kralj-Igli{\v{c}},
  M.~Drab, and N.~S. Gov.
\newblock Theoretical study of vesicle shapes driven by coupling curved
  proteins and active cytoskeletal forces.
\newblock {\em Soft Matter}, 15(26):5319--5330, 2019.

\bibitem{sadhu2021modelling}
R.~K. Sadhu, S.~Peni{\v{c}}, A.~Igli{\v{c}}, and N.~S. Gov.
\newblock Modelling cellular spreading and emergence of motility in the
  presence of curved membrane proteins and active cytoskeleton forces.
\newblock {\em The European Physical Journal Plus}, 136(5):495, 2021.

\end{thebibliography}

\end{document}


\newcommand{\be}{\begin{equation}}
\newcommand{\ee}{\end{equation}}
\newcommand{\bea}{\begin{eqnarray}}
\newcommand{\eea}{\end{eqnarray}}
\newcommand{\nn}{\nonumber}

\title{Supplementary material: Modelling how curved active proteins and shear flow pattern cellular shape and motility}

\author{Shubhadeep Sadhukhan}
\email{shubhadeep.sadhukhan@weizmann.ac.il}
\affiliation{%
 Department of Chemical and Biological Physics, Weizmann Institute of Science, Rehovot, Israel
}
\author{Samo Peni\v{c}}
\email{samo.penic@fe.uni-lj.si}
\affiliation{%
Laboratory of Physics, Faculty of Electrical Engineering, University of Ljubljana, Ljubljana, Slovenia
}
\author{Ale\v{s} Igli\v{c}}
\email{ales.iglic@fe.uni-lj.si}
\affiliation{%
Laboratory of Physics, Faculty of Electrical Engineering, University of Ljubljana, Ljubljana, Slovenia
}
\affiliation{%
Laboratory of Clinical Biophysics, Faculty of Medicine, University of Ljubljana, Ljubljana, Slovenia
}
\author{Nir Gov}%
\email{nir.gov@weizmann.ac.il}
\affiliation{%
 Department of Chemical and Biological Physics, Weizmann Institute of Science, Rehovot, Israel
}
\date{\today}
\maketitle
\renewcommand{\thefigure}{S-\arabic{figure}}
\renewcommand{\thesection}{S-\arabic{section}}
\renewcommand{\theequation}{S-\arabic{equation}}

\section{Simulation details}
The time evolution of the vesicle in our MC simulations consists of \cite{Fonari2019} (1) vertex movement, and (2) bond flip. In a vertex movement step, a vertex is chosen and attempts to move to a position randomly chosen within a sphere of radius $s$ drawn around the vertex position. We set $s=0.15$ in units of $l_{\rm min}$. In a bond flip movement, a bond is chosen common to two neighbouring triangles of the triangulated surface. Two such neighbouring triangles form a quadrilateral. 
The bond connecting two vertices in the diagonal direction of the quadrilateral is destroyed and recreated between the other two vertices, previously unconnected. 
We use the Metropolis algorithm to update the system---a statistical mechanics simulation method to study complex systems. The new state of the system increases the energy of the system by an amount $\Delta E$ is accepted with the rate ${\rm exp}(-\Delta E/k_BT)$. However, the new state is accepted certainly if such a move decreases the system energy.

In the simulations presented in this paper, we use the model parameters as follows: the total number of vertices $N=1447$, the bending rigidity $\kappa=20~k_BT$, protein-protein interaction strength $w=1~k_BT$. The adhesion interaction length is set to be $l_{\rm min}$. Among $N$ vertices, $N_c$ vertices are occupied by the curved membrane proteins. The spontaneous curvature at all the $N_c$ vertices is set to $c_0=1~l_{\rm min}^{-1}$. The spontaneous curvature for the rest of the vertices is set to zero. The percentage of curved protein vertices $\rho=100N_c/N$ is an important parameter. Throughout this paper, we calculated the velocity of the center of mass in units of $l_{\rm min}$/2000 MC steps whereas all the lengths are measured in units of $l_{\rm min}$.

\section{Modelling of shear force}
\begin{figure}
    \centering
    \includegraphics[scale=0.45]{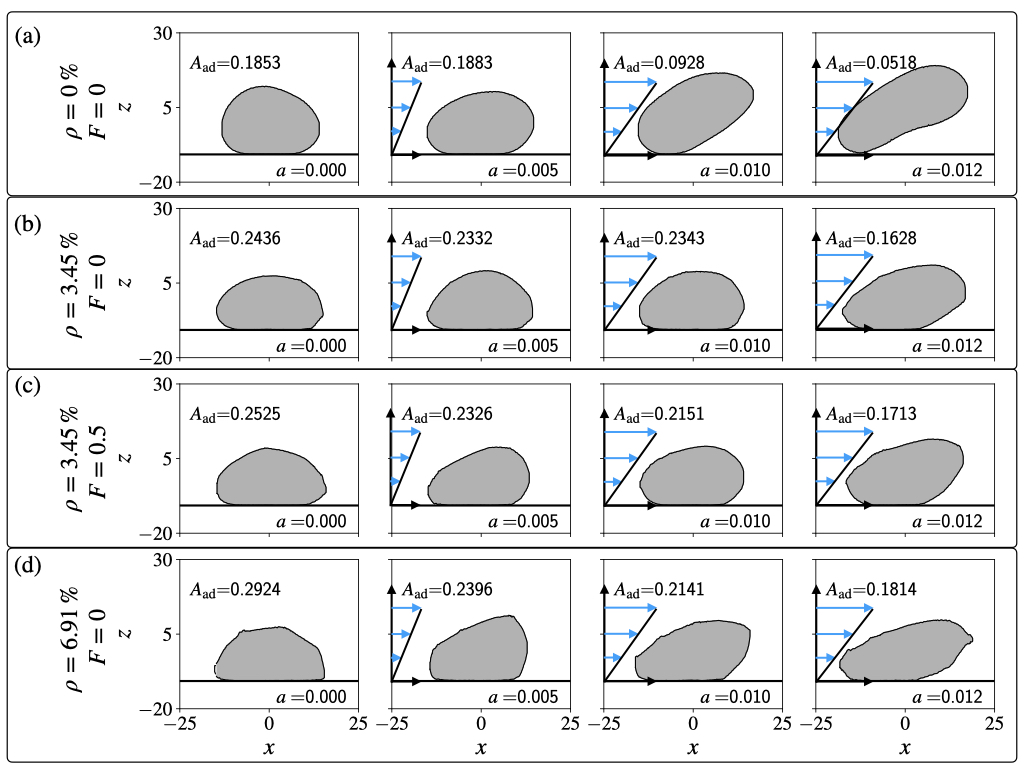}
\caption{Cross-sectional area of the vesicle at the vertical plane that cuts through the center of mass at $y=y_{\rm avg}$, and we applied the shear force along the $\hat{x}$ direction. Each column corresponds to a different shear rate: $a=0,~0.005,~0.01,~0.012 (k_BT/l_{\rm min}^2)$. We set the adhesion strength $E_{\rm ad}=0.5 k_BT$. Four rows are corresponding to the parameter set (a) No protein case: $\rho=0\%,~F=0$; (b) Passive proteins case: $\rho=3.45\%,~F=0$; (c) Active proteins case: $\rho=3.45\%,~F=0.5 k_BT/l_{\rm min}$; (d) Passive proteins case with higher density: $\rho=6.91\%,~F=0$.}
    \label{fig:shear_model}
\end{figure}
We modelled the force due to the shear flow such that the membrane feels a force that is tangential to the membrane surface that is perpendicular to the local normal of the membrane (Fig.1). The direction of the local shear force depends on the global shear flow direction $\hat{v}^{\rm shear}$ and the local normal to the membrane $\hat{n}$. First, we find the direction which is perpendicular to both the flow direction and the local normal at the membrane by a cross product: $\hat{v}^{\rm shear}\times\hat{n}$. The flow should pass normally through the plane formed by the local normal $\hat{n}$ and the direction $\hat{v}^{\rm shear}\times\hat{n}$. Therefore, we modelled the direction of the force due to shear flow as:  
 $\hat{n}\times(\hat{v}^{\rm shear}\times\hat{n})$. We always set the strength of the shear force by the shear rate parameter $a$, and the distance to the adhesive surface $z_i-z_{\rm ad}$. The magnitude of the force due to shear on a vertex positioned at $\overrightarrow{r_i}=(x_i,~y_i,~z_i)$ is $F^{\rm shear}=a(z_i-z_{\rm ad})$. 

We simulated four cases with four different shear rates as shown in Fig~\ref{fig:shear_model}. Four columns correspond to a shear rate $a=0,~0.005,~0.01,~0.012 (k_BT/l_{\rm min}^2)$. Four rows correspond to different cases with different parameter sets: (a) No proteins, (b) protein percentage $\rho=3.45\%$ and no active force $F=0$, (c) protein percentage $\rho=3.45\%$ and active force $F=0.5 k_BT/l_{\rm min}$, (d) protein percentage $\rho=6.91\%$ and no active force $F=0$.
We kept the location of the centre of mass in the $x-y$ plane fixed for all the simulations discussed in this section. 

We find how shear forces affect the shape of the vesicle by plotting the cross-section of the vesicles in the $x$-$z$ plane as shear flow is applied in the $x$ direction (Fig.~\ref{fig:shear_model}). The protein-free shapes that we obtain are very similar to the shapes found in ~\cite{Cantat1999a}, including the tendency of the vesicles to lift and eventually detach as the shear flow strength is increased.
We calculated the magnitude of the adhered surface $A_{\rm ad}$ for each case shown in Fig.~\ref{fig:shear_model} (given in each panel), which we find to be decreasing with the shear rate $a$. We find that the vesicle spreads more when the protein percentage $\rho$ is higher for any fixed shear rate. We also find that the vesicle with active curved proteins spreads more than the vesicle with the passive proteins \cite{sadhu2021modelling}, for a fixed protein percentage $\rho$ and the shear rate $a$. 

\section{The body axis}
The body axis of an unstructured vesicle is defined as its most elongated internal axis, as follows.
The vesicle has $N$ vertices. The $i$th vertex has the position vector $\overrightarrow{r_i}=(x_i, ~y_i, ~z_i)$. The centre of mass of the vesicle is given by the position vector $\overrightarrow{CM}=(CM_x,~CM_y, ~CM_z)$. We find the matrix $G$ given by,
\begin{equation}
     G=\begin{bmatrix}
   G_{xx} & G_{xy}\\
  G_{yx} & G_{yy}
   \end{bmatrix},
\end{equation}
where, $G_{xx}=\sum_i{(x_i-CM_x)^2}, G_{xy}=G_{yx}=\sum_i{(x_i-CM_x)(y_i-CM_y)}, G_{yy}=\sum_i{(y_i-CM_y)^2}$. Next, we found the eigenvector that corresponds to the largest eigenvalue, and it gives us the body axis along which the vesicle is elongated.

\begin{figure}
    \centering
    \includegraphics[scale=0.45]{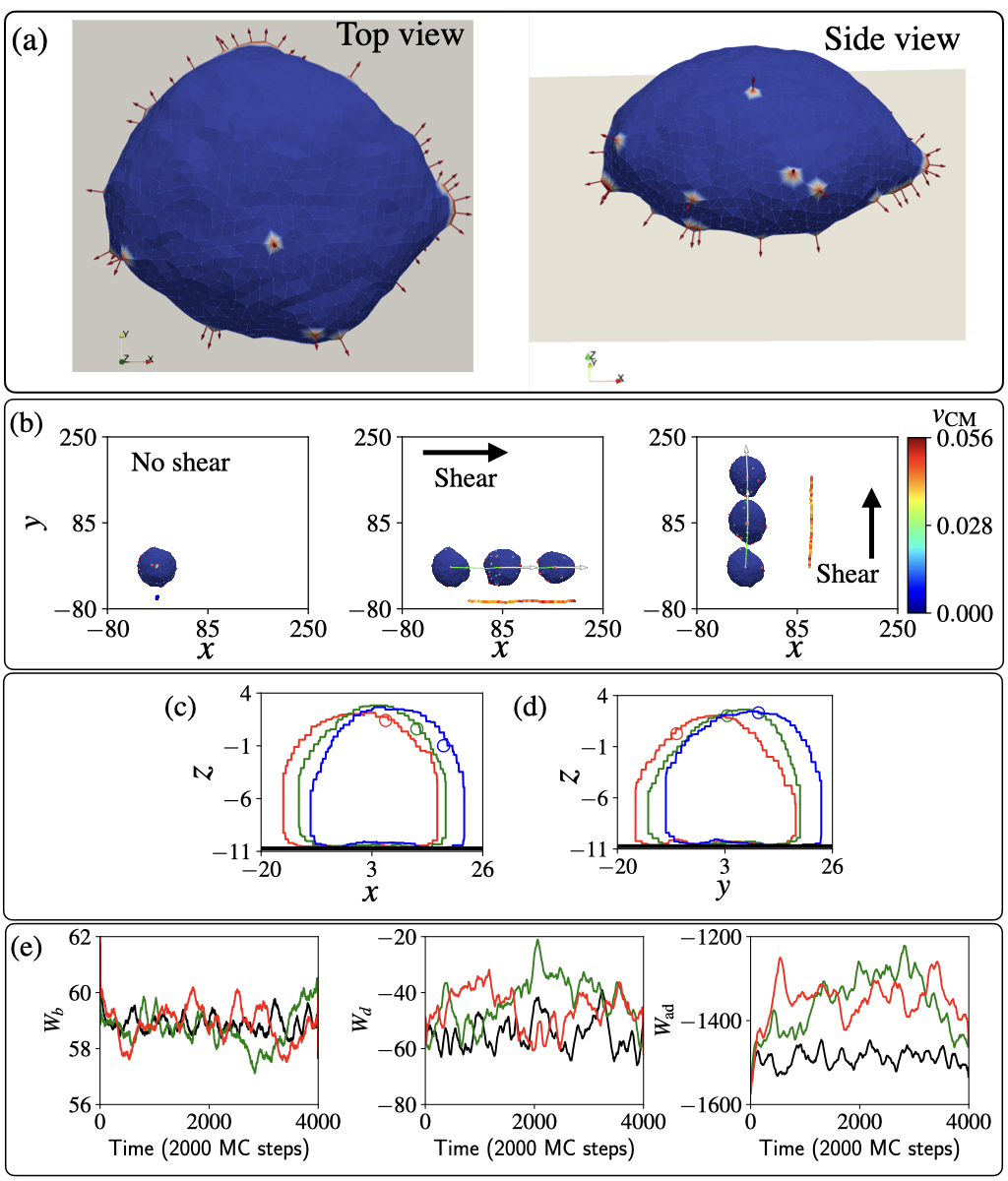}
\caption{Nonpolar vesicle: (a) A weakly spread vesicle on a weakly adhesive surface and low active force \cite{sadhu2021modelling} $F=0.5 k_BT/l_{\rm min},~E_{\rm ad}=1k_BT,rho=3.45\%$. (b) Trajectories of the vesicle for three different shear conditions (no shear, shear in $x$ direction, and shear in $y$ direction) in lime colour, in three respective columns. We set the shear rate $a=0.01 k_BT/l_{\rm min}^2$. A shifted trajectory is shown on each panel with a color code that shows the velocity of the centre of mass of the vesicle. (c) Cross-sections of the vesicle on $x$-$z$ plane when the shear is in the $x$ direction at Time =0, 75, and 100 (unit of 2000 MC steps) in red, green, and blue colours. A particular vertex is encircled on the cross sections. (d) Cross-sections of the vesicle on $y$-$z$ plane when the shear is in the $y$ direction at Time =0, 75, and 100 (unit of 2000 MC steps) in red, green, and blue colours. A particular vertex is encircled on the cross sections. (e) Time evolution of bending energy, protein-protein binding energy, and adhesion energy in three respective panels. Black, green and red lines correspond to the cases of no shear, shear in $x$ direction, and shear in $y$ direction respectively.}
    \label{fig:nonpolar2}
\end{figure}
\section{Non-polar shaped vesicles under shear}
There are several different kinds of non-motile vesicles that emerge in our model \cite{sadhu2021modelling}, other than the two-arc-shaped vesicle presented in the main text (Fig.3). 

\begin{figure}
    \centering
    \includegraphics[scale=0.45]{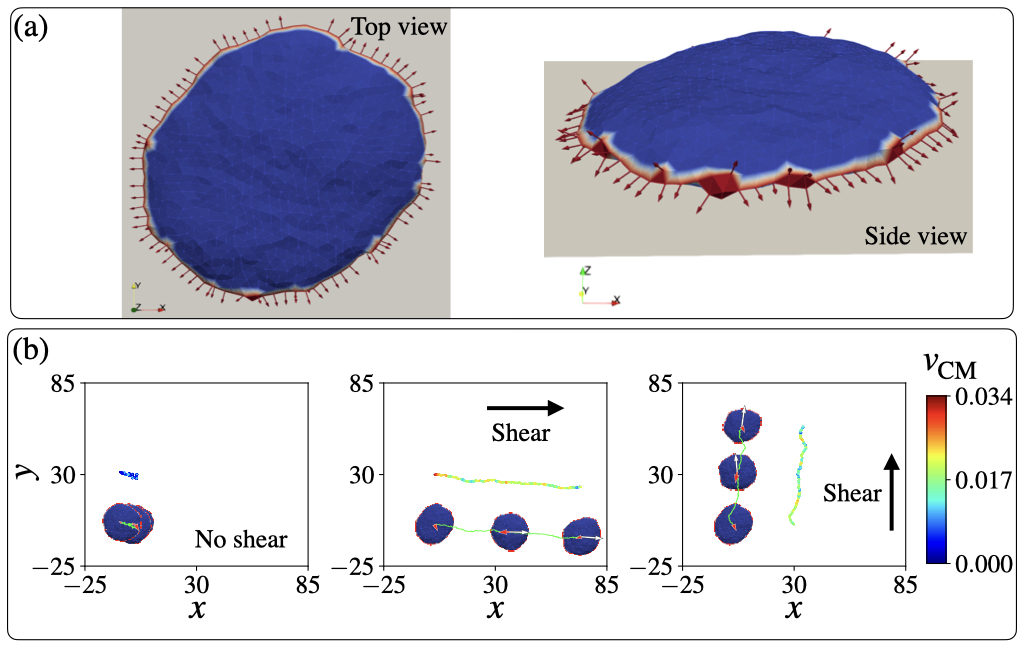}
    \caption{Pancake-shaped vesicle: (a) We set $F=2k_BT/l_{\rm min},~E_{\rm ad}=0.5k_BT, ~\rho=6.91\%$ to get a pancake-shaped vesicle \cite{sadhu2021modelling} (b) The trajectories (lime solid lines) of a pancake-shaped vesicle under different shear conditions in three different columns. We set the shear rate $a=0.01k_BT/l_{\rm min}^2$ for the second and third panels. The active force and the total force are shown in red and white arrows respectively in each column. The velocity of the centre of mass is shown with a solid shifted trajectory with colours mapped to its velocity $v_{\rm CM}$ for each case.}
    \label{fig:pancake}
\end{figure}

\subsection{Weakly spread vesicle under shear}

We can get non-motile non-spreading vesicles when the active force and adhesion strength are low ($F=0.5~k_BT/l_{\rm min}$, $E_{\rm ad}=1 k_BT$) and protein concentration is low $\rho=3.45\%$ (Fig~\ref{fig:nonpolar2}a). The proteins are scattered randomly on the membrane, therefore they can not produce a significant and persistent force in a particular direction. Hence, the vesicle remains at the same position and fluctuates under thermal noise when no shear flow is present, as shown in the first column of Fig~\ref{fig:nonpolar2}(b). We set the shear rate $a=0.01~(k_BT/l_{\rm min}^2)$ for all the sheared cases in this section. The vesicle rolls with the flow as shown in Fig~\ref{fig:nonpolar2}(b), where shear is in the $x$ and $y$ direction respectively. 

Next, we shed some light on the question of how the vesicle is moving with the shear flow. We plot the cross-section of the vesicles on the vertical plane through their center of mass, and the direction of shear flow in Figs.~\ref{fig:nonpolar2}(c,d). By following the motion of a particular vertex on the cross-section we see that the vesicle moves with the flow, performing a mixture of sliding and rolling.

Next, we plot the bending energy $W_b$, protein-protein binding energy $W_d$, and adhesion energy $W_{\rm ad}$ of the system in Fig~\ref{fig:nonpolar2}(e). It is very clear that adhesion is stronger in the absence of shear. It indicates the tendency of shear to cause a lifting force and de-adhesion of the vesicle (as we showed also in Fig.\ref{fig:shear_model}) is the dominant effect.

\subsection{Pancake-shaped vesicle under shear}
If we increase the protein density in such a way that the proteins can form a closed circular cluster around the rim of the vesicle, we can get another type of non-motile vesicle, i.e. a pancake-like shape (Fig.~\ref{fig:pancake}a). In order to get a pancake-shaped vesicle we set a higher active force strength $F=2k_BT/l_{\rm min}$, adhesion strength $E_{\rm ad}=0.5 k_BT$ and most importantly a higher protein percentage $\rho=6.91\%$.

As the active force is equal in every direction due to the formation of the circular rim cluster (Fig.~\ref{fig:pancake}a), the net force on the vesicle is zero. Therefore, without any other force, this shape is non-motile on the adhesive substrate, as shown in Fig.~\ref{fig:pancake}(b). However, the addition of the shear force breaks the force balance, and the vesicle slides with the flow. We studied such pancake-shaped vesicles under the shear flow along $x$ and $y$ directions respectively. The trajectories of the centre of mass of the vesicles under such shear flow show that the vesicle is moving with the flow as shown in Fig.~\ref{fig:pancake}(b). We do not observe significant polarization of the vesicle with the flow, so that the internal active forces do not contribute to the shear-induced migration.

\section{Movies}
\textbf{Movie 1: Polar vesicle}
Polar vesicle is obtained using parameters $E_{\rm ad}=3 k_BT$, $F=2 k_BT~l_{\rm min}^{-1}$, and $\rho=3.45\%$. Four panels for four different shear conditions. When the shear flow is absent, the motile vesicle moves in the $x$ direction persistently. It shows the U-turn of the polar vesicle when $\hat{v}^{\rm shear}=\hat{x}$. Protein aggregate rotates and the vesicle moves against the shear when $\hat{v}^{\rm shear}=\hat{y}$.

\textbf{Movie 2: Two-arc vesicle}
Non-motile two-arc-shaped vesicle is obtained $E_{\rm ad}=1 k_BT$, $F=3 k_BT~l_{\rm min}^{-1}$ and $\rho=3.45\%$. Three panels for three different shear conditions. Without shear, the vesicle retains its position under thermal fluctuation. In presence of shear, it moves with the flow.

\textbf{Movie 3: Non-motile vesicle}
Non-motile weakly spreading vesicle is obtained $E_{\rm ad}=1 k_BT$, $F=0.5 k_BT~l_{\rm min}^{-1}$ and $\rho=3.45\%$. Three panels for three different shear conditions. The vesicle moves with the shear flow, however, it remains at the same position without shear flow because the vesicle is non-motile.

\textbf{Movie 4: Pancake vesicle}
Non-motile pancake-shaped vesicle is obtained $E_{\rm ad}=0.5 k_BT$, $F=2 k_BT~l_{\rm min}^{-1}$ and $\rho=6.91\%$. Three panels for three different shear conditions. The vesicle slides with the shear flow whereas no shear condition leads the vesicle to remain at the initial position without moving significantly.

\bibliographystyle{abbrv}
\bibliography{shubhadeep_biophys}